\documentstyle[aps,prl,psfig,epsfig]{revtex}
\newcommand{\be}{\begin{equation}}
\newcommand{\ee}{\end{equation}}
\newcommand{\bea}{\begin{eqnarray}}
\newcommand{\eea}{\end{eqnarray}}

\tightenlines
\begin{document}
\draft

\title{Strong electron correlations in cobalt valence tautomers}

\author{M.X. LaBute, R.V. Kulkarni, R.G. Endres, D.L. Cox}

\address{
Department of Physics, University of California, Davis, CA 95616}
\twocolumn[\hsize\textwidth\columnwidth\hsize\csname
@twocolumnfalse\endcsname

\date{\today}

\maketitle

\begin{abstract}
We have examined cobalt based valence tautomer molecules such as
Co(SQ)$_2$(phen) using density functional theory (DFT) and variational
configuration interaction (VCI) approaches based upon a model
Hamiltonian.  Our DFT results extend earlier work by finding a reduced
total energy gap (order 0.6 eV) between high temperature and low
temperature states when we fully relax the coordinates (relative to
experimental ones). Furthermore we demonstrate that charge transfer picture
based upon formal valence arguments succeeds qualitatively while failing
quantitatively due to strong covalency between the Co 3$d$ orbitals
and ligand $p$ orbitals.  With the VCI approach, we argue that the
high temperature, high spin phase is strongly mixed valent, with about
30\% admixture of Co(III) into the predominantly Co(II) ground state.
We confirm this mixed valence through a fit to the XANES spectra.
Moreover, the strong electron correlations of the mixed valent phase
provide an energy lowering of about 0.2-0.3 eV of the high temperature
phase relative to the low temperature one.  Finally, we use the 
domain model to account for the extraordinarily large entropy and
enthalpy values associated with the transition.
\end{abstract}

\draft
\vskip1.5pc]

\narrowtext

\section{Introduction}

Transition metal complexes with redox active ligands have been the
subject of extensive studies in recent years.  Of particular interest
are the valence tautomers based upon semiquinone and catecholate
groups ligated to transition metal ions
\cite{pierpont01,shultz01}. These molecules have intrinsic interest as
candidates for mechanical and magnetic switching devices activated by
temperature, pressure, or irradiation. They also present coordination
environments and structural and magnetic conformation reminiscent, at
least, of the local properties in certain allosteric
metalloproteins\cite{allosteric}.

In this paper, we focus on the cobalt based valence tautomers of
generic (high temperature) form Co(SQ)$_2$L$_p$, where SQ represents a
semiquinone ligand complex with formal charge -1 and spin 1/2, and
L$_p$ represents a neutral redox passive counterligand such as phen,
py$_2$X (X=O,Se,S,Te), tetramethylmethylenediamine (tmmda), or
tetramethylethylenediamine (tmeda). When cooled in frozen organic
solvents such as toluene, or in molecular solid form (typically under
pressure), these molecules undergo a transition, in a narrow
temperature regime (typically between 100 and 400K), from a high spin,
high volume form, to a low spin, low volume form
\cite{adams93,adams96,jung94,pierpont95,jung97,jung98}.  The moment
changes from order 4-6 $\mu_B$ at high T to 2 $\mu_B$ at low T, while
the bond lengths of the inner coordination sphere typically decrease
by about 10\%, about 0.2$\AA$.
 
Traditionally, it is assumed that there is a charge transfer from Co
to the SQ ligands concomitant with the transition, with the enhanced
ligand field splitting arising from the bond contraction favoring low
spin Co(III).  This picture accounts for the observed susceptibility
data, and is consistent with observed hyperfine splittings, optical
absorption.

However, there remain some puzzles in this accepted description of the
valence tautomers, notably:
\begin{itemize}
\item{(1) On the theoretical side, density functional calculations
\cite{adams97} have so far provided modest support to the above
description, yielding stable high spin states for the high temperature
geometry, and stable low spin states for the low temperature geometry.
However, more Co electron charge is found at low $T$. }
\item{  (2) For the Co(SQ)$_2$(phen) complex, x-ray absorption near
edge spectra (XANES) spectra measured for the Co K-edge \cite{roux96},
there is additional spectral weight at high energies (about 2-3 eV
above the XANES peaks associated with transitions to 3$d$ states) which
has not yet been accounted for.}
\item{ (3) The estimated entropy change $\Delta S$
at the
transition inferred from fits to the effective moment is extraordinarily
large, of order 10-15 $R$ \cite{adams96}, which is 
difficult to account
for from softened vibrational modes at higher $T$.  Concomitantly,
a large enthalpy of order 0.3 eV/molecule is required. Direct
measurements of $\Delta S$ \cite{abakumov93} for molecular solid forms give, 
on the
other hand, a value of about 1.2 $R$, consistent with the spin entropy
change, which gives $\Delta H \approx 0.03$ eV/molecule. }

\end{itemize}

Motivated by these experimental and theoretical puzzles, we have set
about to extend the description of the cobalt valence tautomers driven
in large measure by a desire to assess the role of strong electron
correlations in such systems.  To this end, we have: (i) performed
new, spin polarized, fully relaxed density functional theory
calculations.  These calculations provide an energy lowering of the
high $T$-low $T$ splitting which partly reduces the disagreement
between earlier DFT calculations and experiment.  A Mulliken
population analysis reveals 0.2 {\it fewer} Co 3d electrons for the
high-$T$ form when compared to the low-$T$ form, in marked contrast to
the standard picture. We demonstrate, through an
examination of the projected densities of states, that the arguments based
upon formal oxidation states are qualitatively correct when one
identifies the overall molecular orbitals derived from the nominal
$e_g$ symmetry about the Co site.  (ii) We have developed a model
Hamiltonian for the active electronic states which we solve by a
variational wave function calculation which includes strong electron
correlation effects on the Co site (outside DFT).  This calculation
amounts to configuration interaction with a physically motivated basis
set reduction which has proven useful for analyzing solid state
systems. We find a significant further reduction in the
high-$T$/low-$T$ energy difference, with stronger electron correlation
effects in the high-$T$ phase (which has 0.3 less Co 3d electrons than
the low-$T$ phase). In particular, at high-$T$ the high spin Co(II)
significantly admixes with a high spin Co(III) state in which an SQ
electron antiferromagnetically screens the Co moment.  We show that
this a good quantitative account for the K-edge XANES data discussed
above. (iii) We show that a domain model \cite{sorai74,kahn93} can resolve the
discrepancy between inferred and direct measurements of $\Delta
S,\Delta H$. Specifically, we show that existing experimental data
is inconsistent with the assumption of a random mixture of
high-$T$ and low-$T$ forms of the tautomer and is best explained
by assuming the molecules form clusters of size $\sim$30-50.


We stress that this description of the high-T phase of the valence
tautomers makes them {\it mixed valent} in the sense used in the
physics community to describe lanthanide compounds such as CeSn$_3$ or
SmS, in which the ground state is a quantum superposition of states
with predominantly two to three well defined valences for the rare
earth ion. This mixed valence is distinct from the multi-site mixed
valence of, e.g., the Creutz-Taube molecule. We distinguish this
single site mixed valence from {\it covalence}, which is applicable to
strong molecular orbital admixture at the single particle level.  In
this sense, a mixed valent single ion is intermediate between fully
localized and fully covalent.  Such molecular states have already been
noted, e.g., in the case of cerocene \cite{cerocene}, where formal
valence arguments generate the expectation of diamagnetic, tetravalent
Ce ions, while observations yield temperature independent
paramagnetism.  The Ce ion is mixed valent in this case.

We discuss the DFT results in Sec. II, the many body theory in Sec. III,
the domain model in Sec. IV, and conclusions in Sec. V. 
   
\section{Electronic Structure Calculations}

\subsection{Computational Method}

We carry out the spin-polarized electronic structure calculations
using the {\it ab initio} code SIESTA \cite{ordejon96,portal97}.  The
Kohn-Sham equations are solved with the exchange-correlation
calculated using the generalized-gradient-corrections(GGA)
approximation, in the fully {\it ab initio} version of Perdew, Burke
and Ernzerhof \cite{perdew96}.  We use Troullier-Martins
norm-conserving pseudopotentials \cite{troullier91} in the
Kleinman-Bylander form \cite{kleinman82}. For Co, we use nonlinear
partial-core corrections to account for exchange and correlation
effects in the core region \cite{louie82}. The basis set orbitals are
obtained using the method of Sankey and Niklewski \cite{sankey89}, 
generalized for multiple-$\zeta$ and polarization functions 
\cite{artacho99}. For
the Co atom the basis set consisted of double-$\zeta$ 3d and 4s
orbitals with 4p polarization orbitals. For C,N and O the basis set
was double-$\zeta$ 2s and 2p orbitals whereas for H it was
double-$\zeta$ 1s orbitals.

The initial atomic coordinates are obtained using the experimentally
determined high-temperature and low-temperature geometries for the
phen complex. In both cases, following previous work \cite{adams97},
the tertiary {\it butyl} groups were removed and replaced with H atoms
(at a distance $\sim 1 \AA $ from the C atoms). The supercell
approximation was used and a cubic supercell of dimension 38 Bohr was
taken to ensure sufficient vacuum between neighboring tautomer
molecules. The results obtained were checked for convergence by
increasing supercell size to 45 Bohr. Conjugate-gradient relaxation
was performed on the system, with the experimental geometries
providing the set of initial coordinates, to determine the minimum
energy configuration within DFT. The atoms were allowed to relax till
the force on each atom was less than 0.04 eV/$\AA$. The electronic
structure was determined both for the initial geometry and the final
relaxed geometry and the results are presented in the following.

\subsection{Results}


The ground-state energetics obtained by us for the experimental
geometries are in agreement with previous {\it ab initio} calculations
\cite{adams97}. For the high-temperature geometry, the electronic
structure calculation yields high-spin ($\sim 3/2$) Co coupled
ferromagnetically to spin $1/2 $ on each of the ligands. Furthermore
we find that the low-spin configuration cannot be stabilized for the
high-temperature geometry. In contrast, for the low-temperature
geometry, the low-spin ($\sim 0$) configuration is stabilized and the
high-spin configuration is unstable. In the low temperature geometry
there is a net spin $1/2$ for the complex which is located on the
oxygen ligands. Furthermore,  we find a net spin of 0.06 on the Co ion, 
consistent
with the observed hyperfine splitting of 30G or less\cite{adams96},
reduced by an order of magnitude compared to, e.g., metallic Co.
Thus, our results are in agreement with experimental data
which show a transition from a high-spin state to a low-spin state as
the temperature is decreased.

The reason is of course, not complicated: the increased 3$d$-2$p$
hybridization upon contraction increases the hybridization induced
ligand field splitting on the Co site, which disfavors Hund's rule
alignment of the spins (the point charge contribution increases as
well). The entropy cost is offset by the energy gain from increased
ligand field energy.

The ground-state energy difference between the two states {\it without
relaxation from the experimental geometry} is found to be 1.3 eV,
which is in good agreement with the previously obtained value of 1.2
eV. However this energy difference is too high to be consistent with
the room-temperature transition observed in the system. Experimentally
the enthalpy change ($\Delta H$) between the two states is inferred
from fits to the magnetic moment to be $\sim 0.3$ eV/molecule
\cite{adams96} which is much lower than the theoretical value (as
mentioned, the lone direct specific heat measurement suggests $\Delta
H\approx$ 0.03 eV/molecule). However, even this experimentally
inferred value for the enthalpy corresponds to an entropy change of
$\sim 14~R$ which is unphysically large. 
Thus we are faced with a discrepancy between the experimental
results and the energetics obtained using DFT.

This discrepancy can be partially resolved by noting that the
experimentally determined ground-state geometry can differ slightly
from the DFT ground-state geometry. Thus to allow the system to relax
to the DFT ground-state configuration, we performed conjugate-gradient
relaxation starting from the experimental geometries. Relaxing the
geometries resulted in a substantial lowering of the energy difference
between the high-spin and low-spin configurations to 0.6 eV, as
compared to 1.3 eV for the unrelaxed geometries. In this relaxation
process, the bond distances between Co and ligands changes very
little, and thus the relaxed geometry might be argued as a better
comparison point for experiment.  In a previous study for the
unrelaxed experimental geometries, Adams {\it et al} \cite{adams96}
obtained a lowering of the ground-state for the high-spin
configuration by performing a `broken-symmetry' run. However, we have
found that the broken-symmetry configuration is not stable and upon
relaxation the ground-state configuration is given by the
ferromagnetic alignment of the ligand spins to the Co spin.  However,
the ground-state energy difference obtained after relaxation is still
too high to be consistent with the low transition temperature and it
indicates that correlation effects which are not included in DFT play
a significant role in determining the ground-state energies.

Performing the Mulliken population analysis for the relaxed geometries
leads to another discrepancy with the traditional interpretation of
valence tautomerism. In this framework, the transition is described as
a spin crossover phenomenon associated with charge transfer from the
metal to the ligands. However the DFT results indicate that the net
$d$-charge {\it increases} in going from the high-spin geometry
($7.1$) to the low-spin geometry ($7.3$). This should be contrasted
with the traditional picture in which we should expect the net
$d$-charge to be lower by 1 in the low-spin geometry due to charge
transfer to the ligands.

\begin{figure}
\parindent=1.0in
\vspace{0.1cm}
\indent{
\psfig{file=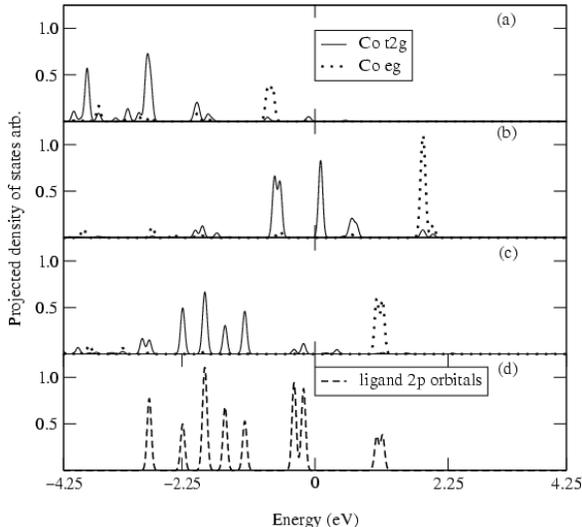,height=3.0in,angle=270}}
\parindent=0.5in
\caption{Partial densities of states (PDOS) for Co-valence tautomers 
in the vicinity of the first ionization level. (a) Co 3d-projected  
PDOS for the majority spin in the high temperature phase. (b) Co 3d- 
projected PDOS for the minority spin in the high temperature phase. 
(c) Co 3d-projected PDOS for majority spin in the high temperature
phase. There is negligible difference in this case between majority
and minority spins. (d) PDOS jointly projected to SQ/CAT 2p and Co 3d 
levels in the low temperature phase.  Note that the formal $e_g$ levels,
for example, are strongly covalently admixed between the ligands and  
the Co ion.}
\end{figure}

A resolution of this discrepancy is reached by analyzing the
transition in terms of molecular orbital theory and studying the
projected density of states (PDOS) for the two geometries. In terms
of molecular orbital theory, the transition can be summarized as
follows. The $d$ orbitals on the metal are split into the $t_{2g}$ and
$e_g$ groups. The metal $e_g$ levels mix with the ligand $\sigma$
orbitals to form molecular orbitals which are bonding and antibonding
in character. The bonding levels are expected to be ligand dominated
and the antibonding levels metal dominated. The $t_{2g}$ levels don't
mix with the $\sigma$ orbitals (ignoring, for simplicity, the mixing
with the ligand $\pi$ orbitals) and the splitting between the $t_{2g}$
levels and the antibonding molecular orbitals (dominated by metal
$e_{g}$ levels) is what corresponds to the crystal-field splitting. In
the high temperature geometry the lower crystal-field splitting
stabilizes the high-spin phase in which the antibonding $e_g$ orbitals
are occupied by two electrons and the $t_{2g}$ levels have 1 hole.  In
the low temperature geometry the $e_{g}$ levels are unoccupied and the
$t_{2g}$ levels are filled and the remaining electron is transferred
to the ligands. Thus we have a crossover from a high-spin
configuration to a low-spin configuration associated with charge
transfer from the metal to the ligands.

A detailed analysis of the PDOS, shown in Fig. 1, for both the
high-spin and low-spin geometries reveals that the above picture is
indeed accurate, and yet it is also consistent with the DFT result
that the net $d$-charge is greater in the low T phase. This is
explained as follows : In the high T phase there is an occupancy of
the $e_{g}-\sigma$ antibonding molecular orbitals of 2 electrons in
the spin-up channel. In the low T phase one of these electrons gets
transferred to the Co $t_{2g}$ level and the other is transferred to
the ligand SQ level. However this does not translate into a lowering
of charge on the Co atom by one. There is significant covalency
between the metal-$e_g$ levels and the ligand-$\sigma$ orbitals.  So
when 2 spin-up electrons are transferred from the $e_g-\sigma$
antibonding orbitals that corresponds to a loss of just 1.4 for the Co
atom and not 2. On the other hand because there is greater overlap in
the low-T phase there is greater metal contribution to the occupied
$e_{g}-\sigma$ bonding orbitals (which are ligand dominated). The     
difference between the metal contributions to the $e_{g}-\sigma$      
bonding orbital is $\sim$ 0.6 (higher in the low T phase).  So the net
metal $e_{g}$ occupancy in going from high T to low T changes by $  
-1.4 + 0.6 = -0.8$ i.e by less than 1 electron and not 2 as the formal
picture indicates. This is more than compensated by the increase in
the occupancy of the $t_{2g}$ level by 1 and hence we see that the net
charge {\it increases} in the low T phase. 

Thus we have seen that the DFT results support the traditional
qualitative picture for valence tautomerism at the same time
clarifying that there is no effective charge transfer in the
system. However there is still a significant discrepancy between the
calculated and observed energy differences between the two states. In
the next section we will explore the role of correlation effects in
reducing this discrepancy.

\section{Variational Many Body Theory}

In this section, we show that electron-electron interaction effects
can provide a further reduction of $\Delta H$ beyond DFT, and that
they provide an explanation for heretofore unexplained spectral
intensity in the high-T XANES spectra.  Moreover, they confirm that
like cerocene, the high-T phase of the valence tautomers is mixed
valent in the sense used within the physics community to describe SmS,
CeSn$_3$ and other intermetallic compounds.

\subsection{Model Hamiltonian}

In order to investigate the possible role for correlation effects in
the cobalt valence tautomers, we wish to consider the simplest model
that possesses the salient features that are commonly accepted as
essential for the combined metal-ligand charge transfer (CT)/spin
crossover of the tautomeric interconversion.  We model this with an
Anderson Impurity Hamiltonian \cite{one} , in which local electron
correlation effects are treated accurately on the Co site, but
neglected on the `metallic' carbon rings.

We have taken the structure of the Co(phen) complex as the physical
basis for our model.  We represent the o-quinone ligands by their
benzene ring skeletons and the oxygen atoms which are nearest neighbor
to the metal site, and neglect octahedral symmetry breaking at the Co
site.  The model system therefore consists of the cobalt atom, the
3$d$-levels split by the cubic field arising from the local six-fold
coordination of the oxygen atoms, the nitrogen atoms of the (N-N)
complex, and the aromatic rings of the semiquinones.

We restrict our model to the electronically active $\pi$ orbitals of
the semiquinone ligands and the 3$d$ $t_{2g}$ and $e_g$ orbitals of the Co ion.
     
The LCAO Hamiltonian using second-quantized notation for
the simple model we have just described is 
\begin{eqnarray}
{\cal H}_{0} = -\sum_{\stackrel{i\alpha,j\beta}{\sigma}}t_{i\alpha
j\beta} c_{i\alpha \sigma}^{\dag}c_{j\beta \sigma}
\end{eqnarray}
where $c_{i\alpha\sigma}^{\dag}$ and $c_{j\beta\sigma}$ are,
respectively the fermion creation and annihilation operators of the
aforementioned orbital set where {\em i} and {\em j} are site indices,
$\alpha$ and $\beta$ are orbital symmetry-adapted labels, and $\sigma$
refers to the spin degeneracy. The {\em t}$_{i\alpha j\beta}$ are
on-site energies for {\em i}$\alpha$ = {\em j}$\beta$ and are
electron-hopping integrals for {\em i}$\alpha \neq$ {\em j}$\beta.$
The on-site 3$d$ energies are
\begin{eqnarray}
\epsilon_{d\gamma\sigma} = \epsilon_{d\sigma} + \Delta\epsilon_{LF}
\end{eqnarray}
where
\begin{eqnarray}
&\Delta\epsilon_{LF}& = -\sum_{i\alpha\sigma}\frac{\mid\langle
 i\alpha\sigma\mid\Delta V\mid
 d\gamma\sigma\rangle\mid^{2}}{\epsilon_{d\sigma} -
 \epsilon_{p}(i\alpha\sigma)} \\ &+&\left\{ \begin{array}{ll} -0.4\Delta_{0} & \mbox{if $\gamma 
= {\em x^{2}-y^{2}, 3z^{2}-r^{2}}$}
 \nonumber \\ 
0.6\Delta_{0} & \mbox{if $\gamma = {\em xy, yz, xz}$}
 \end{array} \right.
\end{eqnarray}
where $\Delta V$ is the potential barrier that an electron hopping
from the Co atom to one of the ligand N or O atoms must tunnel
through, $i$ is the site sum over the 4 oxygens and 2 nitrogen and
$\alpha$ refers to the three 2$p$-orbitals on those sites. $\gamma$ sums
over the degeneracy of irreps of O$_{h}$ and $\epsilon_{d\sigma}$ is
the bare d-electron energy while $\epsilon_{p}$(i$\alpha\sigma$) are
the ligand atom on-site energies. $\Delta_{0}$ is the electrostatic
contribution to the ligand field (LF). $\epsilon_{d\gamma\sigma}$ then
plays the role of the on-site 3$d$-orbital energy in the model.

We include electron-electron interactions only for the Co 3$d$
electrons.  These are certainly the most significant owing to the more
localized character of the states, and experience with transition
metal oxide solids show this to be a good starting assumption.
Including also the spin-orbit coupling, we thus add to ${\cal H}_0$
\begin{eqnarray}
{\cal H}_{d} &=& U_{d}\sum_{\gamma\sigma\ne\gamma^{\prime}\sigma^{\prime}}
n_{\gamma\sigma}n_{\gamma^{\prime}\sigma^{\prime}}+J_{H}\sum_{\stackrel{\gamma > 
\gamma^{\prime}}{\sigma > \sigma^{\prime}}}\vec{s}_{\gamma\sigma}
\cdot\vec{s}_
{\gamma^{\prime}\sigma^{\prime}}\nonumber \\
&+& \sum_{i=1}^{N_{d}}\xi(\vec{r}_{i})\vec{l}_{i}
\cdot\vec{s}_{i}
\end{eqnarray}
where {\em U$_{d}$} is the direct Coulomb integral and {\em J$_{H}$}
$<$ 0 is the ferromagnetic Hunds' rule exchange coupling. The
occupancy operator is defined as {\em n$_{\gamma\sigma}$ =
d$_{\gamma\sigma}^{\dag}$d$_{\gamma\sigma}$} where {\em
d$_{\gamma\sigma}^{\dag}$} creates an electron in the 3$d$-orbital
$\gamma$ with spin $\sigma$.  $\vec{l}_{i}$ is the orbital angular
momentum of the ith d-electron and $\vec{s}_{i}$ is the one-electron
spin. $\xi(\vec{r}_{i})$ is the free-ion parameter.

\indent 
For parameters, we make use of the tabulated on-site energies
and also the distance parameterizations of the hopping integrals of
Harrison \cite{two} . The hopping matrix elements are written as
linear combinations of these integrals using Slater-Koster theory
\cite{three}.  {\em U$_{d}$} and {\em J$_{H}$} are treated as model
parameters.  We are currently working to constrain these parameters
from information gained through ab-initio methods.  However, we may
obtain reasonable estimates by fitting the results of the model
calculation to X-ray absorption data. This places constraints
on the charge transfer gap 
and also {\em U$_{dc}$-U$_{d}$}, which is the difference
between the Coulomb integrals describing the core hole-3$d$ electron  
and 3$d$-3$d$ interactions, respectively. 

The high and low temperature expressions of the charge
transfer gap are given by
\begin{eqnarray}
&\Delta_{CT}^{\mathrm ht}& \simeq \epsilon_{d}-0.4\Delta_{0}-
\frac{3}{2}J_{H}+6U_{d}-\epsilon_{L} \nonumber \\
&{\mathrm and}& \\
&\Delta_{CT}^{\mathrm lt}& \simeq \epsilon_{d}+0.6\Delta_{0}
+6U_{d}-\epsilon_{L}+{\mathrm 1.74 eV} \nonumber
\end{eqnarray}
where $\epsilon_{L}$ refers to the energy of an electron
localized in a ligand orbital and 1.74 eV refers to 
the hybridization contribution to the ligand field splitting
in the low-temperature geometry.

\indent 
Thus, the Hamiltonian is essentially that of a single-impurity
Anderson Model, with the Co ion providing the localized states and
the SQ ligands playing the role of the `metal'.  Such a heuristic
mapping to this model allows us to pursue a method of solution that we
discuss in some detail in the next subsection.

\subsection{Variational Wave Function}

We proceed to solve for the ground state by performing a variational
configuration interaction (VCI) within a restricted basis set
consisting of the lowest energy single-Slater determinants. This
amounts to diagonalizing the Hamiltonian of Eqns.(1) and (4) within this basis
of multi-electron wavefunctions.  These determinants are built from
the valence orbital set of our model.
   
For the `impurity' model described above, the variational
wavefunction method (VCI) has been shown to successfully describe
transition metal or rare earth ions embedded both in metals and in
metal oxides \cite{four,five}.  In the present molecular context we
regard it as a physically motivated basis set reduction: we have
`divided' to conquer by emphasizing first the strongest source of
electron-electron correlations, expanding in the hybridization about
the atomic limit for the Co ion. This said, the approach yields
results which are {\it non-perturbative} in $V$, and systematically
controlled by two handles: (i) the spin+orbital degeneracy (the lowest
order results within a restricted Hilbert space become exact for large
degeneracy), with particle-hole excitations suppressed by inverse
powers of the degeneracy relative to the starting state, and (ii) the
presence of excitation gaps which suppress contributions from
particle-hole excitations about the leading order configurations.
  
The composition of the valence electron ground state will indicate the
significance of correlations. The repulsive nature of the Coulomb
interaction will tend to drive $d$-electrons off the metal site. This
should manifest as coherent quantum mechanical tunneling between
states of $n_{d}$ electrons and states belonging to the $n_{d}$-1
configuration (valence fluctuations).  If such effects are
insignificant, we should expect the type of single-determinant
ground state that is treated very well within Hartree-Fock theory.

The construction of the many-body basis is briefly outlined. The states
are Slater determinants that will consist of tensor products
of cobalt states and ligand states, i.e. 
$\mid\Gamma\beta SM_{S};L_{\alpha}\rangle$
where $\Gamma$ is the crystal point-group irrep, $\beta$ is its
degeneracy, {\em S} is the total spin, {\em M$_{S}$} is the spin
degeneracy.  {\em L$_{\alpha}$} is the ligand state label which is
used only if the $\alpha^{th}$ orbital is occupied. Fig. 2 depicts 
the states we consider in the variational ansatz.
We restrict our Hilbert space to the two low lying states containing 6
or 7 Co 3$d$ electrons; strong Coulomb repulsion legislates against
other configurations.

Magnetic susceptibility and x-ray absorption data, as well as
first-principles calculations, have suggested the relevance of three
closely- lying multiplets for the 3$d$-electrons: the $^{4}$T$_{1}$ of
the high-spin (h.s.) 3$d^{7}$ ({\em S} = $\frac{3}{2}$), the $^{2}$E
of low-spin (l.s.) 3$d^{7}$ ({\em S} = $\frac{1}{2}$), and the
$^{1}$A$_{1}$ of l.s.-3$d^{6}$ ({\em S} = 0). In the spin sector, we
consider only the stretched state of maximal {\em M$_{S}$} within the
{\em S} manifold. These would be expected to be the lowest-lying due
to the Hund's rule energy.

We also include an intermediate spin sector of states ({\em S} = 1,
3$d^{6}$($^{3}$T$_{1}$ and $^{3}$T$_{2}$) which couple to the h.s. and
l.s.  states through the spin-orbit interaction and to l.s.-3$d^{7}$
by CT.

We assume that the SQ/CAT eigenstates of {\em H$_{p}$} are adequately
treated within the framework of H\"{u}ckel theory.  This provides the
standard approximate method of dealing with $\pi^{\ast}$ electrons. We
employ these methods to find the unoccupied states of the two
o-quinone ligands and also their decomposition onto the constituent
atomic orbital basis.  Once these states have been obtained, Eqns.(1) and (4)
may be diagonalized within the truncated basis of tensor product
states.

\begin{figure}[tb]
\epsfysize=6.0cm
\centerline{\epsffile{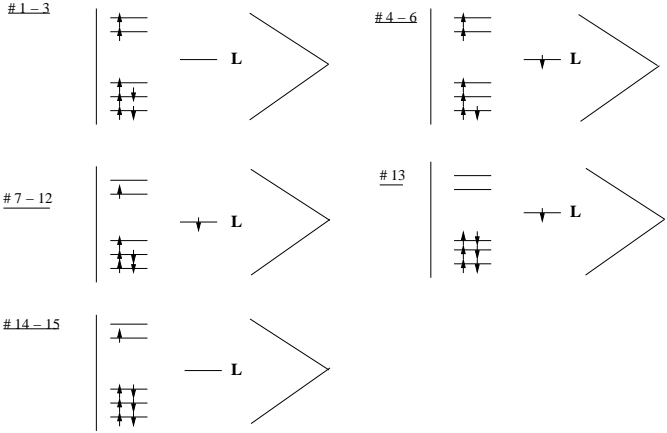}}
\vspace{0.1cm}
\caption{Single-Slater determinants which comprise the many-body basis.
States (1-3) are $^{4}$T$_{1}$ of h.s.-3$d^{7}$, (4-6) belong to $^{5}$T$_{2}$ of h.s.-3$d^{6}$,
(7-12) are $^{3}$T$_{1}$, $^{3}$T$_{2}$ of ($S$=1) 3$d^{6}$, (13) is $^{1}$A$_{1}$ which is
l.s.-3$d^{6}$, and (14-15) is $^{2}$E of l.s.-3$d^{7}$. $L$ represents the
doubly-degenerate spin-down LUMO of the quinone ligands.}
\end{figure}

\subsection{Results}

The Hamiltonian is diagonalized in both the high-T and low-T
geometries using the optimized coordinates from the fully relaxed DFT
runs.  Fig. 3 is a histogram depicting how the quantum weight is
distributed among the determinants of our variational ansatz
ground state in both the high and low temperature phases. The high-T
phase suggests a definite non-trivial role for correlations with a
mixed valent state of 69$\%$ h.s.-3$d^{7}$, 28$\%$ h.s.-3$d^{6}$ with
an electron delocalized over the catechol ligand, and the remainder
($\sim$3$\%$) in l.s.-3$d^{7}$.  The single-determinant description
works quite well in the low-T phase with 98.5$\%$ of the weight
residing in the l.s.-3$d^{6}$ state, and ($\sim$1.5$\%$) residing in
the intermediate spin 3$d^{6}$ state. The small mixing between
low-spin and higher-spin within a single configuration is due to the
spin-orbit coupling. Therefore our low-T phase is exactly consistent
with previous results of both ab initio calculations and experiment.

\vspace{0.0cm}
\begin{figure}
\epsfysize=6.0cm
\centerline{\epsffile{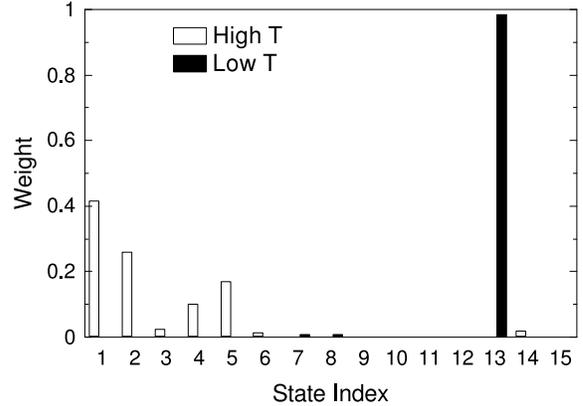}}
\caption{Quantum weights of single determinants in Co(phen) ground state wavefunction.
The vertical axis measures $\mid \alpha_{i}\mid^{2}$ where $\alpha_{i}$ are the variational coefficients.
The model parameters correspond to $\Delta_{CT}^{\mathrm ht}$ = -0.05 eV
and $\Delta_{CT}^{\mathrm lt}$ = 0.42 eV.}
\end{figure}

\indent 
The most surprising aspect of these results is the resonance
state in the high-T phase featuring large admixture with the
h.s.-3$d^{6}$ state. There is a concomitant lowering of the energy
such that the high-T phase is stabilized with respect to the low-T
phase by $\sim$0.4 eV.  This entanglement of states would not be
resolved by a mean-field theory like Hartree-Fock or DFT which
both utilize a single-determinant groundstate composed of effective
single-particle molecular orbitals.

\indent 
An estimate should be made of the contribution of these
correlations to the high-T ground state energy. We can
separate out the contribution to the resonant energy lowering due to
covalency by finding the energy difference between single determinant
h.s.-3$d^{7}$ and the high-T ground state of our calculation. This will be
the energy omitted by DFT calculations.  For completeness, we include
the possibility of the local electron tunneling to all excited states
of the o-quinone ligands. This leads to an extra energy of $\sim$0.2 eV that
needs to be subtracted from the enthalpy difference between high- and
low-T. This further reduces the 0.6 eV obtained from our relaxed DFT
calculations.
\indent 
The small magnitude of the gap $\Delta_{CT}$ seems to be
indicative of the system's proximity to a charge-transfer instability
within our model.  This also justifies the high degree of
inter-configurational admixture in the high-T
phase. $\Delta_{CT}^{\mathrm ht}$ is nearly zero , slightly stabilized on 
the
$n_{d}$ = 7 side. $\Delta_{CT}^{\mathrm lt}$ is larger and has changed 
signs,
so $n_{d}$ = 6 is stable. It will be seen in the next section that
near-edge X-ray absorption data mandates some constraints as to where
the high-T and low-T phases must reside relative to the instability.

\subsection{ XANES calculation}

We find that data taken in X-ray absorption studies near the cobalt
K-edge \cite{roux96} corroborates the existence of the
h.s.-3$d^{6}$/3$d^{7}$ mixed valence in the high-T state. We first
discuss features of the experimental spectra. The on-set of the 1$s$-4$p$
absorption edge is $\sim$7712.2 eV. The region of immediate interest to
us is the low-intensity pre-edge interpreted as being composed of
quadrupolar-allowed 1$s$-3$d$ transitions. (Actually, the Co ions lack
inversion symmetry, though if restricted to just nearest neighbors
there is an approximate mirror symmetry forbidding mixing between 4$p$
and 3$d$ states on site.  The symmetry breaking by the SQ rings will
yield an admixture of 4$p$ and 3$d$ which is likely to dominate the
direct quadrupole matrix elements.) Data taken here provides a means
to probe the nominal valence of the cobalt $d$-orbitals. As the
temperature is reduced from T$_{amb}$, three main structures become
apparent. The lowest energy feature is a shoulder at 7708 eV that
loses spectral weight as the temperature is
reduced.  In the high-T phase, there is an intense peak at 7709.6 eV
(which shifts up in energy by $\sim$0.3 eV at lower temperatures) that
gains $\sim$56$\%$ in height. The third structure is a broad shoulder
at $\sim$7712.3 eV that all but completely disappears at low
temperatures. While the first two structures have been assigned the
1$s$-3$d$ {\em t$_{2g}$} and 1$s$-3$d$ {\em e$_{g}$} transitions, this latter
feature has yet to be identified.

We have used our model as the basis to calculate a theoretical
XANES. It has been shown that the core electrons in transition metal
and rare earth compounds do not efficiently screen the 1$s$-core
hole left behind after the photoemission process. Thus the energies of
the 3$d$-valence orbitals are significantly renormalized by Coulomb
interaction with the core-hole. Treated perturbatively this amounts to
a down-shift in energy for these orbitals.  It has also been shown
that in order to obtain good agreement with experimental data, a term
must be added to the Hamiltonian that accounts for this core
hole-valence electron interaction \cite{seven},
\begin{eqnarray}
\delta{\em H} = -U_{dc}\sum_{\sigma}(1-s_{\sigma}^{\dag}s_{\sigma})
\sum_{\gamma}d_{\gamma\sigma}^{\dag}d_{\gamma\sigma}
\end{eqnarray}
so our Hamiltonian in Eqns.(1) and (4) becomes {\em H$^{\prime}$}
\begin{eqnarray}
{\em H^{\prime}} = {\em H} + \delta{\em H}
\end{eqnarray}
where {\em s$^{\dag}_{\sigma}$} is the 1$s$-core electron creation
operator, {\em U$_{dc}$} is the 1$s$-3$d$ Coulomb integral. The expression
we use to calculate the X-ray absorption coefficient {\em I} for the
1$s$-3$d$ transition is the following which is similar to a weighted
many-body density of states
\begin{eqnarray}
I(\omega) = \sum_{\alpha^{\prime}}\mid\langle\alpha^{\prime}\mid\hat{T}\mid
\Psi_{0}\rangle\mid^{2}\delta(\omega-E_{\alpha^{\prime}}+E_{0})
\end{eqnarray}
where $\omega$ is the incident photon energy, {\em
E$_{\alpha^{\prime}}$} and $\mid\alpha^{\prime}\rangle$ are the
eigenstates of the Hamiltonian {\em H$^{\prime}$} which is the same
system described by {\em H} in the presence of the core-hole
perturbation. {\em E$_{0}$} and $\mid\Psi_{0}\rangle$ are the
unperturbed ground state energy and wavefunction of our model
calculation. $\hat{T}$ is the transition operator
\begin{eqnarray}
\hat{T} = W\sum_{\gamma\sigma}d_{\gamma\sigma}^{\dag}s_{\sigma}
\end{eqnarray}
where we have neglected the orbital angular momentum dependence in {\em W}, the
single-electron 1$s$-3$d$ transition matrix element.

We now discuss our results.  Fig. 4(a) shows the spectra directly
resulting from the model we have developed in the previous
sections. The experimental XANES data has been superimposed over our
results. Fig 4(b) depicts the result of a single-determinant
l.s.-3$d^{7}$ low-T ground state for the purposes of comparison; since
the tautomers are on the edge of the charge transfer instability, it
might be possible by suitable selection of the redox passive ligand to
access this low-spin Co(II) state.  We have not estimated the
broadening generated by electronic coupling to vibrational degrees of
freedom and other contributing factors to finite lifetime
effects. This limits our comparison to peak positions and relative
weights.


We first discuss the high-T result. Fig 4(a) features four main
temperature dependent peaks that may be directly related to the XAS
data. We first note that our
calculations reproduce the relative peak heights quite well. The peak
at 7708.03 eV matches the shoulder found in that region and is the
result of a down-spin 1$s$ electron transferring into the {\em
t$_{2g}$}. The shoulder feature has been previously assigned this
transition. The peak at 7709.6 eV is the consequence of the transfer
of a down-spin electron into an {\em e$_{g}$} level, also consistent
with previous assignment.  The energy distance between these two
structures ($\sim$1.60 eV) is approximately the value of the
ligand-field splitting which we would expect on the basis of the
assigned transitions. There are two very closely-spaced peaks at
7710.79 and 7710.8 eV. These are transitions of down-spin electrons
into the {\em t$_{2g}$} shell of 3$d^{7}$ final states.  Going up in
in energy by an amount equal to the ligand-field, we find a cluster
of low-intensity peaks in the interval 7712.32 eV - 7712.42 eV that
correspond to transfers into the {\em e$_{g}$} levels of the 3$d^{7}$
final states. These latter features can be identified with the
spectral weight centered about 7712.3 eV. This assignment is
consistent with the disappearance of these peaks at low-T. To our
knowledge, this is the first attempt to interpret these higher-energy
satellite structures. This feature is thus unique to the type of
highly admixed 3$d^{7}$/3$d^{6}$ ground-state we are proposing for
high-T.  It can now be seen that the distance between the peaks at
$\sim$7709.6 eV and the cluster centered about 7712.4 eV (and also the
distance between the peak at 7708.03 and the peaks at $\sim$7710.8 eV)
of $\sim$2.5 eV provides a constraint for the value of {\em
U$_{dc}$-U$_{d}$}.
\vspace{0.0cm}
\begin{figure}
\epsfysize=9.0cm
\centerline{\epsffile{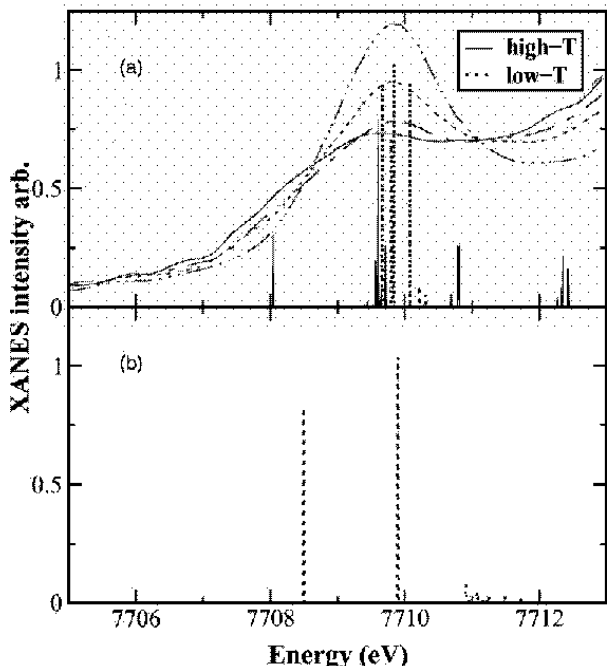}}
\vspace{0.0cm}
\caption{ (a) Model XANES calculation shown with experimental results
for $\Delta_{CT}^{\mathrm ht}$ = -0.05 eV,
$\Delta_{CT}^{\mathrm lt}$ = 0.416 eV, {\em E$_{shift}$} = -1.05 eV,
and {\em U$_{dc}$-U$_{d}$} = 2.36 eV. Experimental curves show change
in XANES with temperature. The solid line corresponds to the high-T  
phase and the (--- $\cdot\cdot\cdot$ ---) line is the low-T phase. 
(b) Low-T XANES calculation resulting from l.s.-3d$^{7}$ groundstate.
$\Delta_{CT}^{\mathrm lt}$ = -0.004 eV, {\em E$_{shift}$} = 0.737 eV, 
and {\em U$_{dc}$-U$_{d}$} = 2.43 eV}
\end{figure}

\indent 
We have introduced an energy shift between the high- and low-T
spectra, {\em E$_{shift}$}. The pragmatic intent is to fit the
data. The underlying physical reasoning is twofold. We have used
identical model parameters for both the high-T and low-T calculations,
allowing only the distance dependence of the hopping integrals to
change. We would expect some renormalization of some of these
parameters due to effects like screening.  More significantly, we have
omitted some terms from our simple model Hamiltonian that would be
quite different in the two geometries, e.g. cohesion energies, off-site 
Coulomb interactions, etc.  Presumably,
these would all have to be incorporated in such a shift.  The low-T
spectra in Fig 4(a) consists of 4 main peak structures concentrated in
the region 7709.66 - 7710.8 eV. These are all associated with
transferring a 1s-electron into the empty {\em e$_{g}$} shell of the
l.s.-3$d^{6}$ state. The heights and location of these agree quite
well with the single broadened peak of experiment.  The weights have
not been altered relative to the high-T phase, giving very good
consistency with experiment.

\indent 
Fig 4(b) depicts a low-T XANES resulting from a l.s.-3$d^{7}$
groundstate. This regime should be investigated since this manifold of
states lies in close energetic proximity to the l.s.-3$d^{6}$
singlet state. As shown, this low-T state can be accessed with only
very slight tuning of the parameters. Such an interconversion would
necessarily lack metal-ligand CT and may be relevant for some species
of tautomer which only exhibits spin-crossover on the metal
site. While the central peak at 7709.9 eV agrees well with the data,
there is the feature of augmented weight at low energies
($\sim$7708.05 eV) at low-T rather than the suppression seen in
experiment. Since the absorption data seem to lack the twin peak
structure (resulting from Hund's rule exchange splitting of up- and down- spin
electron transitions to the singly-occupied {\em e$_{g}$} shell), this
seems to be an untenable possibility for the low-T groundstate, at
least with current choices for redox passive ligands.

\section{Domain Model}


 As noted in the introduction, another problem with the traditional
interpretation of valence tautomerism is the extraordinarily large
entropy change (10-15 R) associated with the transition. However, it
should be noted that this entropy change is inferred only indirectly
from measurements of the magnetic susceptibility and optical
absorption. On the the other hand, measurements of the specific heat
as a function of temperature done by Abakumov {\it et al}
\cite{abakumov93}, which constitute a more direct measurement of the
entropy, give a much lower entropy value ($\sim$ 1 R). This
discrepancy leads us to examine the set of assumptions used to infer
the entropy change from the susceptibility data. These can be
summarized as follows.

At any given temperature $T$, the system is taken to be a
two component mixture of 
the high-spin and low-spin forms of the molecule. The fraction of atoms 
in the high-spin form ($f$) is determined thermodynamically by minimizing the 
molar Gibbs free energy ($G$) which is given by
\begin{equation}
G = fG_{hs} +(1-f)G_{ls} - RT [f{\mathrm ln}(f) + (1-f){\mathrm ln}(1-f)]
\end{equation}
where the last term is the entropy of mixing for a random
two-component mixture. Minimizing the above equation with respect to
$f$ then gives us the functional form for the variation of $f$ with
$T$ which used to infer the enthalpy ($\Delta H$) and entropy ($\Delta
S$) change from the experimental data.

However the above derivation makes the assumption that the system can
be described as a {\it random} two-component mixture. If there is
clustering present in the system, then the above equations have to be
modified. The simplest generalization which accounts for this is the
domain model \cite{sorai74,kahn93}.  In this model, clustering is accounted
for by considering domains with $n$ number of molecules per domain
($n$ is like a mean-field parameter corresponding to the average size
of the domain, with $n=1$ corresponding to the random two-component
mixture). In this case the entropy of mixing is now given by
\begin{equation}
S_{mix} = - R/n~[f{\mathrm ln}(f) + (1-f){\mathrm ln}(1-f) ]
\end{equation}
Correspondingly the variation of $f$ with temperature is 
\begin{equation}
f = 1 / [ 1 + e^{\frac{n \Delta H}{R}(1/T - 1/T_{c})}]
\end{equation}

The key point is that the values inferred from experiment have used
the above form with $n=1$ giving unphysically large values for $\Delta
H$ and $\Delta S$. However if there is clustering ($ n > 1$) then the
inferred value of $\Delta H$ (and correspondingly $\Delta S$) will be
lowered by a factor of $n$ leading to more reasonable values. This
leads us to the question: How can we infer what the appropriate value
of $n$ is?  According to the domain model the answer can be obtained
from specific heat measurements which should indicate a peak in the
specific heat ($C_{p}$) centered around the transition temperature
$T_{c}$. The `jump' in the specific heat is given by $C_{p}(T_{c}) -
\frac{1}{2} (C_{ls}(T_c) + C_{hs}(T_c))$ where
$C_{ls}(T_c)(C_{hs}(T_c))$ is obtained by linear extrapolation at
$T=T_c$ of the low(high) temperature specific heat values. The number
of molecules per domain $n$ is then related to this jump by
\begin{equation}
n = \frac{ 4~R~T_{c}^2}{ (\Delta H)^2 } [ C_{p}(T_{c}) -  \frac{1}{2} 
     (C_{ls}(T_c) + C_{hs}(T_c)) ]
\end{equation}

It should be noted that the specific heat measurements will provide a
direct measurement of of $\Delta S$ (and thereby $\Delta H$) and
within the domain model this can be used to estimate $n$. For the
tautomers, only one set of experiments done so far have measured the
specific heat. From the data given by Abakumov {\it et al} for the bpy
complex, we make the following estimates : The `jump' in $C_p$
corresponds to $\sim$ 150 J/mol~K which gives a change in entropy of $ \Delta
S \sim 1.2~ R$.  Using these values and the equation for $n$ in the
domain model we estimate $n \approx 50$. Thus the data for the
specific heat shows evidence for clustering in the tautomers and is
inconsistent with the original assumption of a random two-component
mixture.  The latter assumption leads to unphysically high inferred
values of $\Delta H$ and $\Delta S$ and we propose that measurements
of the specific heat can be used to resolve this discrepancy and to
self-consistently check for the applicability of the domain model to
the system.

\section{Conclusions}

In conclusion, we have extended the theory of cobalt valence tautomers
to resolve some of the puzzles surrounding various data.  In
particular, we have shown that using fully relaxed density functional
theory solutions goes halfway towards solving the mystery of large
enthalpy values inferred from fits to optical absorption and magnetic
susceptibility data.  However, the relaxed DFT result of 0.6 eV is
still off by a factor of 20 from the sole direct measurement of the
enthalpy found in the tautomer literature\cite{abakumov93}.  We have
also shown that DFT provides little evidence for {\it net} charge transfer 
off 
the Co ion to the SQ ligands (indeed, the 3$d$ weight increases modestly 
in the low-T phase!). On the other hand, a careful examination of the
PDOS shows that the qualitative picture emerging from the formal
valence arguments is still correct: the orbitals with the largest
admixture of 3$d-e_g$ symmetry are depopulated in going through the
transition.

Next, we showed that a variational configuration interaction approach
could produce a further lowering of the high-T enthalpy relative to
the low-T entropy by producing a 30\% admixture of Co(II) and Co(III)
configuration.  At the lowest variational order, we estimate an
0.2-0.3 eV lowering of energy due to electron correlation effects
beyond DFT, which brings us within an order of magnitude of the directly
observed enthalpy.  This high-T state is intrinsically ``CI'': no
unitary transformation in the single particle orbital space can remove
the multiconfiguration character.  It is, in fact, a mixed valent
state associated with the Co ion and the more extended and weakly
correlated states of the SQ rings.  We have also shown that this
high-T state, when proper accounting is made for the core-hole
interaction of the 3$d$ electrons, can explain the previously observed
spectral weight shift well above the dominant 1$s$-3$d$ peak for high
temperatures.  The weight is ascribed to transitions to the admixed
high spin Co(III) configuration.

Finally, we have proposed a domain model to account for the large
$\Delta S,\Delta H$ values inferred from susceptibility and optical
absorption data (as compared to direct measurement).  In this view,
the transition corresponds to a broadened first order phase transition
of molecular clusters of order 20-60. In the dilute solvent, it might
be possible to observe such clustering with elastic light scattering
experiments. Although the heat capacity measurements in the
interesting temperature range are difficult, we certainly would
encourage further experiments on tautomers.

We hope to apply this combined DFT and VCI approach to other tautomer
systems and to analyses of the closely related phenomena observed in
the prosthetic complexes of metalloproteins and metalloenzymes.

{\it Acknowledgements}.  We would like to thank P. Ordej\'on, E. Artacho,
D. S\'anchez-Portal and J. M. Soler for providing us with their {\it ab initio}
code SIESTA. We acknowledge useful discussions with
R.R.P. Singh, A. Shreve, and R. Weht.  The work at Davis was supported by
the U.S. Department of Energy, Office of Basic Energy Sciences,
Division of Materials Research, and by a seed grant from the Materials
Research Institute of Lawrence Livermore Laboratories. This research
also received support from an NSF IGERT "Nanomaterials in the
Environment, Agriculture, and Technology".

\end{document}